\documentclass[useAMS,usenatbib]{mn2e}
\usepackage{graphicx}
\usepackage{amsmath}
\usepackage{amssymb}
\newcommand  \acc     {\ifmmode {\rm km\,s}^{-2} \else km\,s$^{-2}$\fi}

\newcommand  \ergs     {\ifmmode {\rm ergs\,s}^{-1} \else ergs s$^{-1}$\fi}
\newcommand  \ergcms   {\ifmmode {\rm erg~cm}^{-2}\,{\rm s}^{-1}
                        \else erg~cm$^{-2}$\,s$^{-1}$\fi}
\newcommand  \ergcmsA  {\ifmmode{\rm erg\,cm}^{-2}\,{\rm s}^{-1}\,{\rm\AA}^{-1}
                        \else ergs\,cm$^{-2}$\,s$^{-1}$\,\AA$^{-1}$\fi}
\newcommand  \ergcmsHz {\ifmmode{\rm ergs\,cm}^{-2}\,{\rm s}^{-1}\,{\rm Hz}^{-1}
                        \else ergs\,cm$^{-2}$\,s$^{-1}$\,Hz$^{-1}$\fi}
\newcommand  \phcms    {\ifmmode {\rm ph\,cm}^{-2}\,{\rm s}^{-1}
                        \else ph\,cm$^{-2}$\,s$^{-1}$\fi}
\newcommand  \phcmsA   {\ifmmode {\rm ph\,cm}^{-2}\,{\rm s}^{-1}\,{\rm\AA}^{-1}
                        \else ph\,cm$^{-2}$\,s$^{-1}$\,\AA$^{-1}$\fi}

\newcommand\aj{{AJ}}%
\newcommand\araa{{ARA\&A}}%
\newcommand\apj{{ApJ}}%
\newcommand\apjl{{ApJ}}%
\newcommand\apjs{{ApJS}}%
\newcommand\aap{{A\&A}}%
\newcommand\aaps{{A\&AS}}%
\newcommand\mnras{{MNRAS}}%
\newcommand\pasp{{PASP}}%
\newcommand\nat{{Nature}}%
\title{
A Search for the Progenitors of Two Type-Ia Supernovae in NGC 1316}

\author[Dan Maoz and Filippo Mannucci]
{Dan~Maoz$^{1,2}$\thanks{E-mail: maoz@wise.tau.ac.il},
Filippo Mannucci$^{3}$\\
$^{1}$School of Physics and Astronomy, 
Tel-Aviv University, Tel-Aviv 69978,
Israel\\
$^{2}$INAF - Osservatorio Astrofisico di Arcetri, Largo Enrico Fermi 5,
Florence 50125, Italy\\
$^{3}$INAF - Istituto di Radioastronomia, Largo Enrico Fermi 5,
Florence 50125, Italy
}
\date{\today}

\begin{document}

\maketitle

\label{firstpage}

\begin{abstract}
Recent evidence of a young progenitor population 
for many Type-Ia SNe (SNe-Ia) raises the possibility that evolved
intermediate-mass progenitor stars may be detected in pre-explosion 
images. NGC~1316, a radio galaxy in the Fornax cluster, is a prolific 
producer of SNe-Ia, with four detected since 1980.
We analyze Hubble Space Telescope (HST) 
pre-explosion images of the sites of two  of the   
SNe-Ia that exploded in this galaxy, SN2006dd (a normal Type-Ia) and 
SN2006mr (likely a subluminous, 1991bg-like, SN-Ia).
Astrometric positions are obtained from 
optical and near-IR ground-based images of the events.
We find no candidate point sources at either location, and set upper
limits on the flux in $B$, $V$, and $I$ from any such
progenitors. We also estimate the
amount of extinction that could be present, based on
analysis of the surface-brightness inhomogeneities in the
HST images themselves. At the
distance of NGC~1316, the limits correspond to absolute magnitudes of
$\sim -5.5$, $-5.4$, and $-6.0$ mag
in $M_B$, $M_V$, and $M_I$, respectively. 
Comparison to stellar evolution models
argues against the presence at the SN sites, 3 years prior to the explosion,
of normal stars with initial masses $\ga 6M_{\odot}$
at the tip of their asymptotic-giant branch (AGB) evolution, young
post-AGB stars that had initial masses $\ga 4M_{\odot}$, 
and post-red-giant stars of initial masses $\ga 9 M_{\odot}$.    

 \end{abstract}


\section{Introduction}
Although there is general consensus that Type-Ia supernova (SN-Ia)
explosions are the result of a thermonuclear runaway in a degenerate,
approximately Chandrasekhar-mass ($M_{\rm ch}$),
carbon-oxygen stellar core (e.g., Hillebrandt \& Niemeyer 2000;
Badenes et al. 2006; Mazzali et al. 2007),  
the progenitor systems of  SNe-Ia are still
unknown. The two main models leading to a SN-Ia explosion are
the single-degenerate (SD) model, in which a white dwarf (WD)
accretes matter from a close companion until it approaches the
Chandrasekhar limit (Whelan \& Iben 1973; Nomoto 1982), 
and the double-degenerate (DD) model,
involving the merger of two WDs (Iben \& Tutukov 1984; Webbink
1984). However, both scenarios
face theoretical and observational challenges.   
The outcome of a DD merger may be accretion-induced core collapse, rather
than a SN-Ia (e.g., Nomoto \& Iben 1985; 
Saio \& Nomoto 2004, Guerrero et al. 2004), 
although there are 
opposing views that invoke rotation of the stellar surface to prevent
this outcome (e.g., Piersanti et al. 2003).
An observational
search for DD progenitor systems (Napiwotzki et al. 2004;
Nelemans et al. 2005) has turned up, among $\sim 1000$ WDs surveyed, 
few or no potential DD systems
with a total mass exceeding $M_{\rm ch}$ that will merge within a Hubble
time.   

The SD scenario, in turn, has been criticized (e.g., Cassisi et
al. 1998; Piersanti et al. 1999, 2000)
for its assumptions about the existence of {\it ad hoc} mechanisms that
regulate the accretion flow on to the WD (e.g. Hachisu et al. 1996). 
Observationally,
Badenes et al. (2007) have noted the absence, 
in seven nearby SN-Ia remnants, of the
signatures of the strong wind from the accretor that supposedly
stabilizes the accretion flow, 
and permits reaching $M_{\rm ch}$. 
Prieto et al. (2007) have found no evidence for a low-metallicity 
threshold in SN-Ia hosts, in
contrast to the predictions by Kobayashi et al. (1998; see
also Kobayashi \& Nomoto 2008) 
of such a threshold, due to a minimum
metallicity that is required for the wind regulation mechanism 
to be effective. While evidence for circumstellar material, consistent with
expectations from a wind from a red-giant companion, has been found in
one recent normal SN-Ia (Patat et al. 2007), such material is not
observed in other events (Mattila et al. 2005; Simon et al. 2007;
Leonard 2007). 
For the remnant of Tycho's SN, which 
was a type-Ia (Badenes et al. 2006), there have been
conflicting claims about the identification and nature of a remaining companion
star (Ruiz-Lapuente et al. 2004;  Fuhrmann 2005; Ihara et al. 2007).
If, in the end, no companion is found, this would be another problem
for the SD picture. Recent metal
abundance measurements suggest 
a type-Ia explosion also in Kepler's SN remnant 
(e.g., Blair et al. 2007; Reynolds et al. 2007) 
but the
velocity of the remnant away from the Galactic plane,
together with evidence for a circumstellar medium from a massive
progenitor, are problematic the context of a binary progenitor
system (Reynolds et al. 2007).

For core-collapse SNe, which derive from massive ($\ga 8
M_{\odot}$) stars, direct searches of pre-explosion images for
progenitors have had a number of successes -- in the case of SN1987A
(West et al. 1987; White \& Malin 1987),
and several other events with suitable {\it Hubble Space Telescope}
(HST) data (Maund \& Smartt 2005; Li et
al. 2006; Hendry et al. 2006; Gal-Yam et al. 2007). 
In contrast, the expectation, based
on the SD and DD models, that
SNe-Ia explode in old, low-mass, systems, has discouraged such
direct searches for progenitors of SNe-Ia in the HST era. 
While there have been some studies of SN-Ia environments 
using post-explosion HST data (Van Dyk et al. 1999), the reverse has
not been done, largely due to a preconception that the 
progenitors would be undetectable. (Interestingly, Van Dyk et al. 1999
noted young stellar populations in the vicinity of the four SNe-Ia they 
studied, but subtracted them off the images, arguing that the
SNe-Ia were necessarily derived from an old population).

However,
several recent observational developments, in addition to the problems faced by
the popular models, suggest that such a search may be
useful after all.     
Mannucci et al. (2005) have measured stellar-mass-normalized
type-Ia SN rates as a function of galaxy Hubble type and galaxy colors,  
and have found that the SN-Ia
rate in star-forming galaxies traces the star formation 
rate. Such a correlation directly implies that at least some of the
progenitors are young stars.
This study and several subsequent ones (Scannapieco \& Bildsten 2005;
Mannucci et al. 2006; Sullivan et al. 2006; Dilday et al. 2008) 
have demonstrated that there must be a wide distribution
of delay times between star formation and SN explosion, as predicted
by some progenitor models (e.g., 
Greggio \& Renzini, 1983; Greggio 2005).
The spread
in the delay times can be described in terms of two populations
of progenitors:
a ``prompt'' one, which dominates the 
SN-Ia rate in star-forming galaxies and 
has  a typical delay times of less than $\sim 10^8$~yr; 
and ``tardy'' population, 
having a SN-Ia delay time
of $\ga 10^9$~yr (or a large-delay tail to the same progenitor 
population above), needed to produce also the SN-Ia rate measured in old
stellar populations with no current star formation. 
An outstanding puzzle is the possible decrease in the SN-Ia rate at
redshifts $z>1$ measured by Dahlen et al. (2004, 2008), at a cosmic
time when the star-formation rate was an order of magnitude higher
than today. The SN-Ia rate, which would therefore 
have then been dominated by the prompt population, would be
expected to track the star formation rate, as the latter flattens or
continues to rise to high $z$. 
On the other hand, reanalysis of some of these data by Kuznetsova et
al. (2008), and independent high-$z$ rates by Poznanski et al. (2007), 
 have questioned the significance of the 
claimed decrease in SN-Ia rate at high redshift, pointing to the need 
for further observations to resolve the issue.  

In any event, 
the formation of a degenerate carbon stellar core within $10^8$~yr
for a significant fraction of SN-Ia progenitors points to stars with 
zero-age main sequence (ZAMS) masses of $\ga 5 M_\odot$
(e.g. Girardi et al. 2000). 
During their post-main-sequence evolution,
such intermediate-mass stars may reach luminosities high enough to
make them potentially detectable in pre-explosion HST images of nearby
host galaxies. Such 
evolved stars could be present in SN-Ia progenitor systems, either 
in the role of mass donors to the WD (previously produced by a 
more massive primary), where the evolutionary timescale
 of the secondary dictates the onset of the explosion; or in the role 
of single-star SN-Ia progenitors, where degenerate carbon-core
ignition occurs in an evolved star that has somehow lost its hydrogen
envelope (e.g. Tout 2005; Waldman, Yungelson, \& Barkat 2007).

Furthermore, recent measurements of
extragalactic SN-Ia rates at various redshifts have yielded
high rates, especially in star-forming environments.
Maoz (2008) has recently compiled, and compared self-consistently, 
different measurements of various observables related to SN rates,
and used them to estimate the fraction of intermediate-mass close
binaries that explode as SNe-Ia through the prompt
and tardy channels. He shows that the high SN-Ia rates (as well as
other independent observables such as intracluster abundances) indicate that
most or all close, intermediate-mass, binaries must explode as SNe-Ia, 
with, possibly, an excess of SNe-Ia over progenitor systems. This 
conclusion holds despite the uncertainties in the initial mass ranges
that lead to viable WD SN-Ia progenitors in the SD and DD scenarios,
and the uncertainties in the initial parameters of binaries (binarity
fraction, mass ratio distributions, separation distributions). The
conclusion stands
in contrast to detailed SD and DD models that
predict a small exploding 
fraction among the intermediate-mass population. The
independent
observation that a major part of the SN-Ia population derives from a
young population, indicating initial masses in a narrow range 
of $\sim 5-8 M_\odot$ (see
above), further culls the pool of potential progenitors, and makes the
case for an excess of SN-Ia explosions compared to progenitors 
almost unavoidable. A possible solution 
could, again, be evolved, single-star, stripped envelope, SN-Ia
progenitors. Direct searches for such stars in pre-explosion images
of the sites of SNe-Ia in nearby galaxies can test this scenario.

In this paper, we perform such a search in the massive nearby galaxy
NGC~1316, a prolific SN producer that hosted two SN-Ia events in 2006,
by analyzing deep pre-explosion HST images.  After
compiling some facts about this galaxy and its SNe, we
describe the archival HST and ground-based data we have used, and their
analysis for setting flux limits on individual progenitors for each of
the two
SNe. We then discuss the physical limits on SN-Ia progenitors imposed
by these results.   
 
\section{NGC~1316 and its Supernovae}

NGC~1316 (Fornax A) is a giant elliptical radio galaxy in the Fornax
cluster. It is remarkable in many ways. It is extremely luminous
(M$_B$=-22.39)
and massive (log($M/M_\odot$)=11.9, see Mannucci et al., 2005).
A prominent dust lane that cuts across it, with
numerous additional dust features conspicuous throughout the galaxy,
along with some kinematic signatures and analysis of its globular
cluster system, all
suggest that it has undergone a merger  $\sim 3$~Gyr ago
(e.g. Schweizer 1980; Goudfrooij et al. 2004). (A spectacular color
rendition of the galaxy has been produced by the Hubble Heritage team,
{\tt http://hubblesite.org} ). The galaxy hosts a low-luminosity X-ray
 active nucleus (Kim \& Fabbiano 2003), and has a bright pair of radio
lobes separated by $\sim 200$~kpc (Ekers et al. 1983; 
Geldzahler \& Fomalont, 1984).
The distance to NGC~1316 has been measured by Jensen et
al. (2003; $20.0\pm 1.6$~Mpc) using surface-brightness fluctuations; 
by Feldmeier et al. (2007; $17.9\pm 0.9$~Mpc) 
using the planetary nebula luminosity
function; and using the type-Ia SN2006dd (see below) as a distance
indicator with the multicolor light-curve shape method (MLCS, Jha, Riess,
and Kirshner 2007) by P. Garnavich (personal communication; $19.1\pm
1.0$~Mpc, for a Hubble parameter of 70~km~s$^{-1}$~Mpc$^{-1}$). 
We will adopt a distance based on the mean of these three consistent 
measurements, $19\pm 2$~Mpc, corresponding to a distance modulus of
$31.4\pm 0.2$~mag.
The Galactic extinction along the line of sight is low, with $E(B-V) = 0.021$ 
(Schlegel, Finkbeiner, and Davis 1998).

\smallskip

NGC~1316 is the single galaxy that has hosted the largest number of
 discovered and confirmed SN-Ia
events, four since 1980. In view of the facts that routine, CCD-based,
 monitoring
of nearby galaxies has not taken place until recently, and that the high
surface brightness of this galaxy can make more difficult the detection of SNe 
in the bright central regions, it is possible that additional 
SNe have exploded in the galaxy during the past decades, but were
missed. 
For example, based on the data we will 
analyze below, the $V$-band surface brightness within the central $5''$ of the 
nucleus is $13.8-16$~mag asec$^{-2}$. At the distance of the galaxy, this
is comparable to the magnitude of SNe
within a 1~asec$^{-2}$ seeing disk a few weeks past maximum light.
 We summarize
below some information about the four SNe.   

\smallskip

Hamuy et al. (1991) presented spectroscopy and photometry
of SN1980N and SN1981D, showing that both events were normal SNe-Ia
with similar luminosities, reaching $V\approx 12.4$~mag at 
maximum light. Both of these events, at projected
distances of $\sim 10-20$~kpc from the nucleus, occurred outside the
field of view of the HST data analyzed here.   

SN2006dd was discovered by Monard (2006).
While the colors of this SN, analyzed by the method of
Poznanski et al. (2002), initially suggested a Type-II event
(Immler, Miller, \& Brown 2006), spectroscopy by Salvo et al. (2006)
and Morrell et al. (2006) showed it to be a normal SN-Ia.
Like SN1980N and SN1981D, SN2006dd reached $V\approx 12.4$~mag at 
maximum light (see K. Krisciunas's website
at {\tt http://www.nd.edu/$\sim$kkrisciu/sn2006dd.html} for light curves
of this event).

SN2006mr was also discovered by Monard (Monard \& Folatelli 2006),
and classified spectroscopically as a SN-Ia by Phillips et al.
(2006), who noted strong Na I D interstellar absorption (of equivalent
width 3.2~\AA), suggesting ``significant reddening''. Indeed, as opposed to 
the previous three SNe in this galaxy that reached $V\approx 12.4$~mag at 
maximum light (see above), 
SN2006mr reached only $V\approx 14.6$, as measured by the
Carnegie Supernova Project (CSP, Hamuy et al. 2006; see light curves at 
{\tt http://csp1.lco.cl/$\sim$cspuser1/PUB/CSP.html}). However, 
the CSP $B$-band light curve of SN2006mr shows a fading during the
15 days after maximum of $\Delta m_{15}\approx 2$~mag. Such large values
of $\Delta m_{15}$ characterize extremely underluminous SNe-Ia, with 
absolute $B$ magnitudes fainter by $\sim 2$~mag than the mean of the normal 
SN-Ia population (e.g., Altavilla et al. 2004). SN2006mr had a maximum of
$B-V\approx 1.5$ in its
colour evolution about 10 days past $V$-band maximum, very similar 
to the prototype underluminous SNe-Ia 1991bg and 1992K (Hamuy et
al. 1994), but distinct from normal SNe-Ia, that at such times 
have $B-V\approx 0.5$, and reach their reddest phase only about a
month past maximum.   
We conclude that SN2006mr was likely a typical underluminous SN-Ia, rather
than being highly extinguished.

Based on these four events within 27 years, the SN-Ia rate in NGC~1316 is 
much larger than expected in quiescent ellipticals; only 
1.0 SNe, rather than 4, 
are expected in this galaxy during this period, based on
the galaxy's stellar mass and the early-type SN rate measured for
local ellipticals by Mannucci et al. (2005). However, Della Valle et
al. (2005) have found evidence for increased SN rates in radio
galaxies, of which NGC~1316 is a prime example. 
Sharon et al. (2007) showed that cluster galaxies may
have a higher SN-Ia rate than that measured for local early-type
galaxies by Mannucci et al. (2005).
Mannucci et al. (2008) found confirming evidence that cluster ellipticals
have higher SN-Ia rates than field ellipticals, and that this
enhancement is independent of radio loudness. Thus, galaxies that 
are both radio loud and in clusters have SN-Ia rates larger by an
order of magnitude than radio-quiet field ellipticals. 
 Using the SN rate appropriate for its environment and radio-loudness
(Mannucci et al. 2008), 2.1 SNe 
are expected in NGC~1316 in 27 years, while only 0.28 would
be expected in a radio-quiet field elliptical of the same mass.
While keeping in mind the dangers of {\it a posteriori} statistics,
NGC~1316 may thus be an
 individual example of these statistical trends. Although the
dependences of SN-Ia rate on radio loudness and environment still lack
a satisfactory explanation (and also require more observational
confirmation), it is possible that these excesses of SNe-Ia are
somehow related to the recent merger in NGC~1316, and the star
formation that likely ensued. 
If so, some of these SNe could belong to the
prompt channel, further justifying a search for young progenitors
at their pre-explosion sites.  
 
\section{Data and Analysis}

The archival HST data we use, obtained by 
Goudfrooij et al. (2004), were taken with the Advanced
  Camera for Surveys (ACS) in its Wide Field Camera (WFC) mode in 
2003, March 4 and March 7, about
3 years before the two SNe that exploded in this galaxy in 2006. 
Pipeline-reduced (Pavlovsky et al. 2005) images were retrieved from the 
HST archive, consisting of: a F435W (similar to $B$-band) image
with a total exposure time of 1240~s; a F555W (similar to $V$-band) 
image of 6980~s; and two F814W ($I$-band) images having exposures of
2200~s and 2480~s. We aligned and co-added the latter two, for an 
effective total $I$-band exposure time of 4680~s.

\subsection{Astrometry}
To determine accurate positions for SN2006dd and SN2006mr, we
retrieved from the
European Southern Observatory (ESO) archive
images recording these events near maximum light, as follows. 
For SN2006dd, we used near-IR images obtained in the $J$, $H$ and $K$ 
bands with the
SOFI instrument on the New Technology Telescope
 on 2006, August 4, which was 46 days after the discovery of SN2006dd.
The SN is well detected in all bands.
For SN2006mr, we used an $R$-band image obtained with FORS2 on the
Very Large
Telescope on 2006, November 18, which was 13 days after discovery. 
The SN is saturated in the images but its centroid can be determined with a
precision of about 0.5 pixels, i.e., $\approx 0\farcs06$.

The ground-based images were registered with the HST images using
unsaturated stars appearing in both fields. 
The final precisions are $0\farcs3$ for SN2006dd and $0\farcs2$
for SN2006mr.
Table~1 lists the pixel positions of the two SNe in the second
pipeline-reduced HST $I$ image, and their RA and Dec positions, based
on the world coordinate system (WCS) information in the image header. We
note that the WCS data depend on the absolute HST pointing accuracy, 
which is $\approx 1''$. The absolute RA and Dec of the SNe therefore have 
uncertainties of this order, but for the
present study only the (smaller) errors relative to the frame of the HST images
are relevant.

\begin{table}
\begin{tabular}{l|l|l}
\hline
\hline
{}           & {SN2006dd}   &{SN2006mr} \\
\hline
RA(J2000)$^a $&  03:22:41.7&  03:22:43.1\\
Dec(J2000)$^a$&--37:12:13.0&--37:12:29.6\\
pixel$^b$& 2431,2359      &2156, 2750\\
\hline
$B_{\rm lim}^{~~~~c}$&26.0 &25.8\\
$V_{\rm lim}^{~~~~c}$&26.1 &26.0\\
$I_{\rm lim}^{~~~~c}$&25.4 &25.4\\
\hline
$M_{B,{\rm lim}}^{~~~~~~~~~~d}$&$-5.4$&$-5.6$\\
$M_{V,{\rm lim}}^{~~~~~~~~~~d}$&$-5.3$&$-5.4$\\
$M_{I,{\rm lim}}^{~~~~~~~~~~d}$&$-6.0$&$-6.0$\\
\hline
$A_B^{~~~~~~~~e}$&0.4&0.4\\
$A_V^{~~~~~~~~e}$&0.25&0.25\\
$A_I^{~~~~~~~~e}$&0.1&0.1\\
\hline
\end{tabular}

\caption{SN positions and upper limits on flux of progenitor point sources. 
Notes: 
(a) - RA and Dec in the HST WCS frame, accurate to $\approx 1''$;    
(b) - pixel x and y position on the F814W image
j6n201040$\_$drz.fits. Position centroid has an error circle of radius 4
pixels ($0\farcs2$) for 2006mr and 6
pixels ($0\farcs3$) for 2006dd;
(c) - flux limits, in AB magnitudes, based on simulations (see text);  
(d) - absolute magnitude limits based on flux
limits, assuming a distance modulus of 31.4~mag and no extinction
(e) - plausible extinction corrections, in magnitudes,
 based on surface brightness
      inhomogeneities (see text).
    }
\label{table1}
\end{table}

\begin{figure*}
\includegraphics[width=0.85\textwidth]{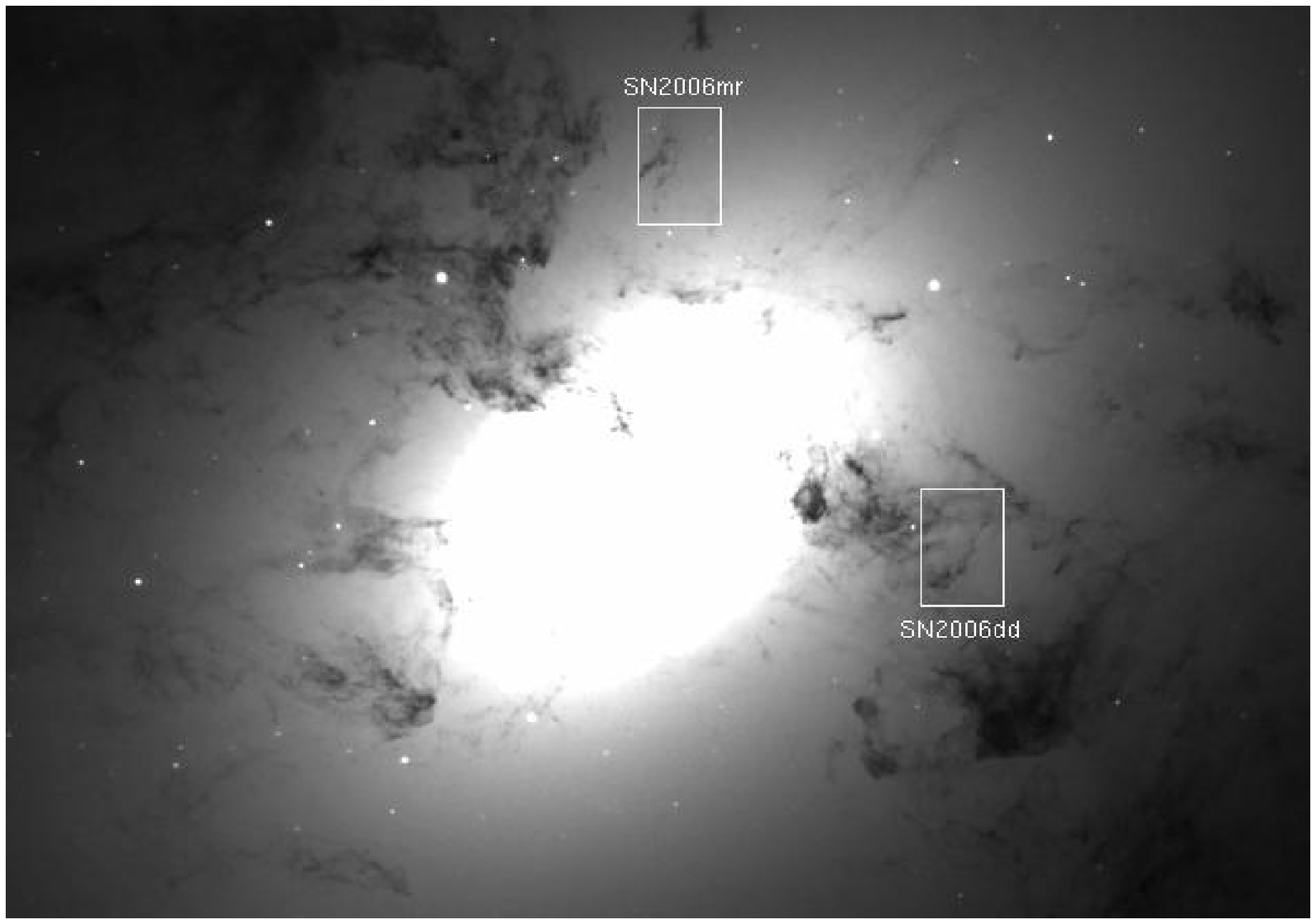}
\end{figure*}
\begin{figure*}
\includegraphics[width=0.85\textwidth]{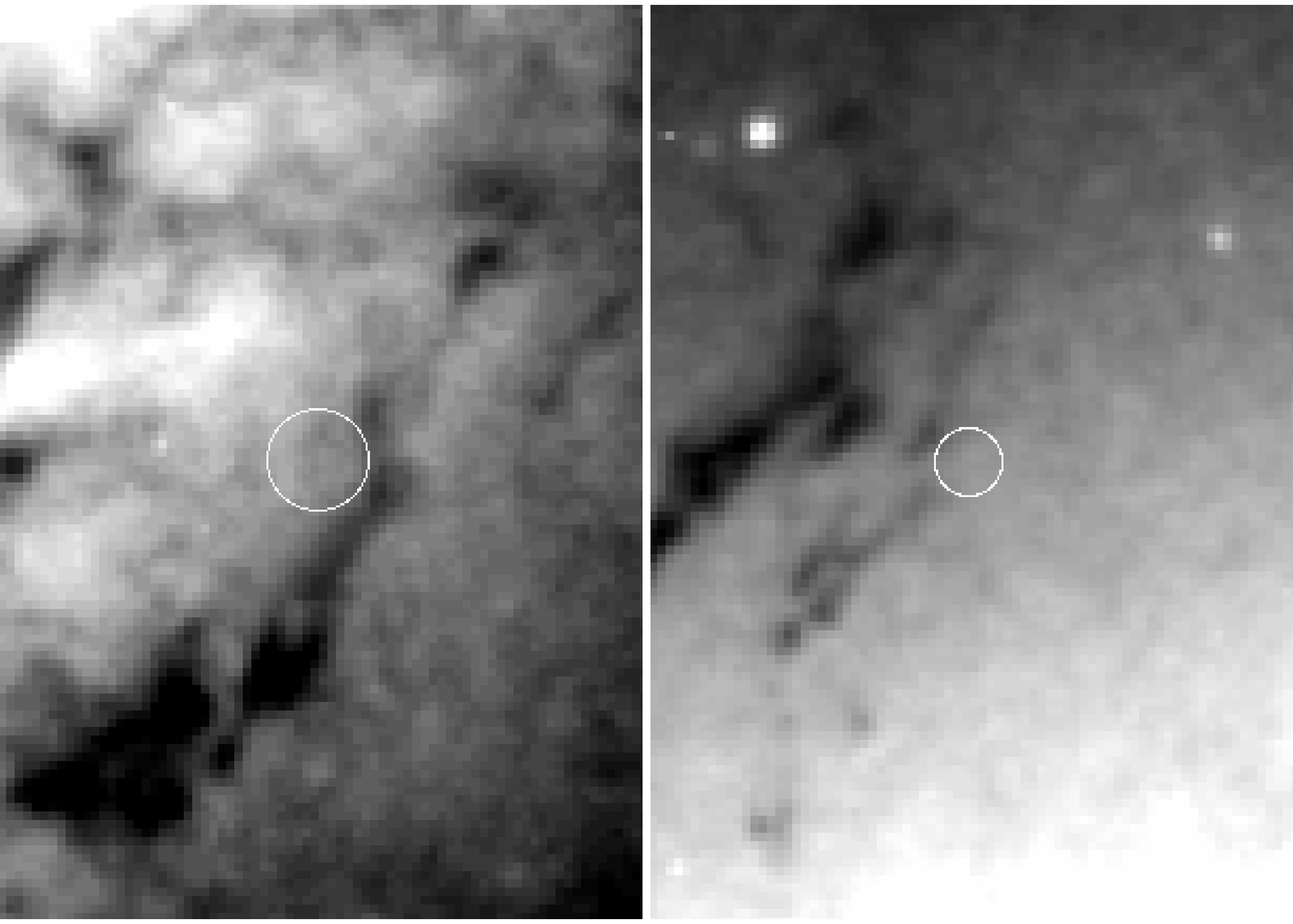}
\caption{{\it Top:} $V$-band HST/ACS image of NGC~1316, showing the locations
of the two SNe. {\it Bottom:} Zoom in on $4''\times 6''$ regions
at the locations of SN2006dd (left) and SN2006mr (right) in the HST
``white light'' image, composed of the sum of the $B$, $V$, and $I$ images.
The $0\farcs2-0\farcs3$ radius error circle at the site of each SN is shown.} 
\label{whitelight}
\end{figure*}

\subsection{Flux Limits}
We visually examined the error-circles of the SNe in the 
$B$, $V$, and $I$ HST images, as well as in a deep ``white light''
image that we formed by registering and co-adding all three bands.
Figure~\ref{whitelight} shows the two sites in the white light image.
We also examined the SN sites as seen in the Hubble Heritage color
image. We find no evidence for any point source above the diffuse 
background at either position. 

\begin{figure*}
\includegraphics[width=0.9\textwidth]{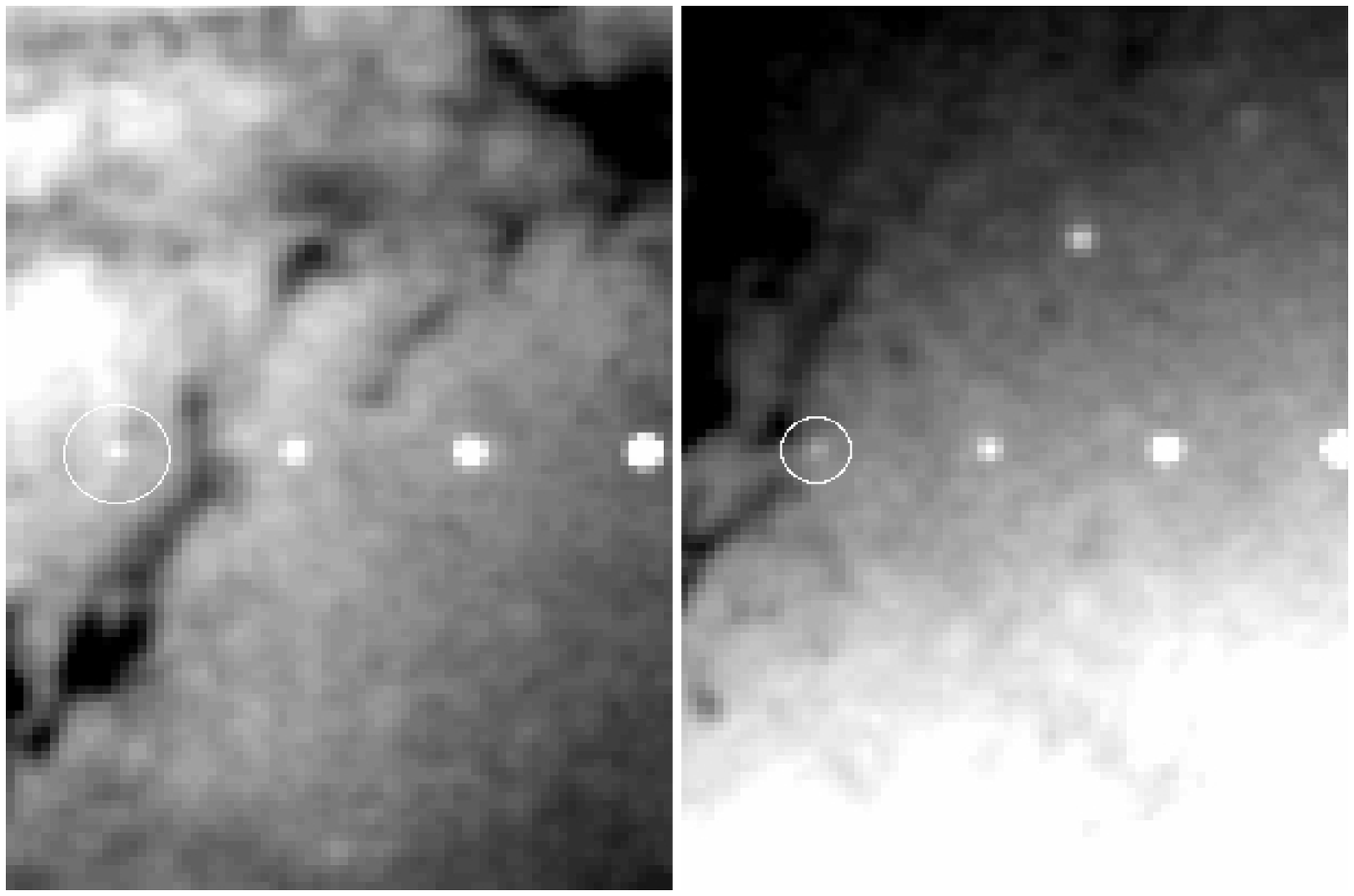}
\caption{Illustration of detection limit simulations at the SN sites. 
Sections, $4''\times 6''$ on a side, of the $V$-band HST/ACS image, 
at the location of SN2006dd (left) and SN2006mr (right),
with a row of simulated point sources added to each.
Every simulated
source is 1 magnitude fainter than the one to its right. The faintest 
detectable ($\sim 3\sigma$) simulated source at each site is shown 
at the center 
of the $0\farcs2-0\farcs3$ radius error circle. 
The $V$ band detection limit is 26.0~mag for SN2006mr, and 26.1~mag
for SN2006dd. The same procedure yields flux limits in $B$ and $I$,
as listed in Table~1}
\label{simV}
\end{figure*}

To set upper limits on the fluxes of
progenitor stars that could be present at the SN sites, we planted
simulated point sources of various magnitudes at the SN positions
in each of the HST images, using the IRAF task {\tt artdata.mkobject}.
The point sources had a Gaussian profile with a
full-width-at-half-maximum of 2.7 ACS pixels in $B$ and $I$, and
2.5 pixels in $V$, matching that of bright, 
isolated, unresolved real sources in the images. Total counts in the
simulated sources were related to fluxes using the PHOTFLAM keyword
in the image headers. Figure~\ref{simV} shows examples of the
simulated data at the SN sites for the $V$ image. Each simulated
source is 1 magnitude fainter than the one to its right. The faintest 
clearly detectable simulated source at each site is shown at the center 
of the error circle. These sources are discernible above the
background fluctuations (which result mainly from the complex
dust patterns) at approximately the $3\sigma$ level.
This $V$ band detection limit is 26.0~mag for SN2006mr, and 26.1~mag
for SN2006dd. 
The same procedure with 
the $B$ and $I$ images gives AB magnitude limits 
of $B=25.8$ and $I=25.4$ for SN2006mr, and  
 $B=26.0$ and $I=25.4$ for SN2006dd. These limits are listed in Table~1.
Finally, assuming the adopted distance modulus of 31.4~mag, 
we give in Table~1 the corresponding absolute
AB magnitude limits on $M_B$, $M_V$, and $M_I$.  

\subsection{Extinction}

As noted in \S2, based on the properties of the SNe themselves, 
it is unlikely that there was high extinction along the sightlines to 
the sites of either SN2006dd or SN2006mr. SN2006dd reached
$V=12.4$~mag, at maximum, very similar to the previous two SNe in this
galaxy, SN1980N and SN1981D. Low extinction is also consistent with
an estimate by P. Garnavich (private communication), using
the MLCS (Jha et al. 2007) method, of a total $V$-band extinction of 
$A_V= 0.24\pm 0.11$~mag toward SN2006dd. As for SN2006mr, as
also noted in \S2, its light curve and color evolution seem
characteristic of an unextinguished, but instrinsically underluminous, 
Type-Ia event.

However, NGC~1316 does have conspicuous dust features, 
raising the possibility that some extinction exists, and could affect
our photometric limits.
 Fortunately, the fact that this galaxy is an
elliptical, with regular iso-brightness contours, permits evaluating 
a reasonable magnitude for this effect based on the images
themselves. To do this, within each
SN error circle, we find the pixel with the lowest flux, and compare 
it to the typical flux in unobscured regions along the same
iso-brightness elliptical contour. This
gives an estimate of the amount of attenuation a progenitor star would
suffer if the dust were in a screen that is completely foreground 
to the diffuse stellar light along the
line of sight, and the progenitor is completely behind the dust
screen. With this procedure,
we find maximum surface-brightness attenuations in $B$, $V$, and $I$
of 0.7, 0.8, and 0.9, respectively, at both SN sites. This then
raises the point-source flux limits by about 0.4~mag, 0.25~mag, and
0.1~mag, in the three bands, respectively, as also listed in Table~1.  

Naturally, less extinction will take place if, in reality,
the star is in front of, or within, the dust clump, rather than behind
it; and the extinction will be underestimated if the dust is behind most of
the diffuse stellar emission, rather than in front of it,
as then the signature of the dust will not be seen. Our procedure thus 
gives a plausible  estimate of the possible attenuation, 
intermediate to these extreme scenarios, although of course we cannot rule out
that these extremes did occur. Another possibility is that a progenitor was
hidden behind a small clump of dust that is unresolved in the HST
images, but which produced arbitrarily high extinction. However, as
discussed above, none of
these extreme scenarios are likely, in view of the colors and
magnitudes of the events themselves.

\section{Discussion}

In this section, we now examine, based on stellar evolution models,
which stellar progenitors present 3 years prior
to the explosion can be ruled out by our limits. 
Metallicity that is close to Solar has been measured in this galaxy, 
both for the stars Goudfrooij et al. (2001a,b) and for the gas 
Kim \& Fabbiano (2003).

\begin{figure*}
\includegraphics[width=0.85\textwidth]{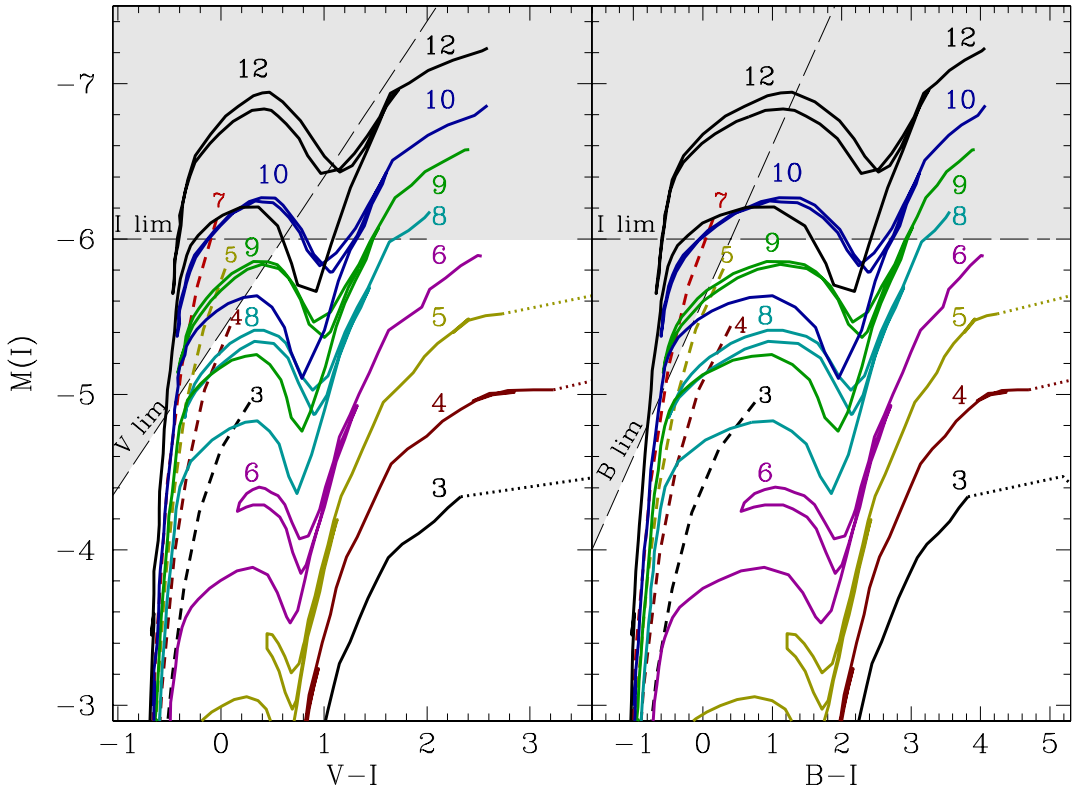}
\caption{Model stellar evolution tracks from ZAMS to AGB (solid
  curves, from Girardi et
  al. 2000, for $3-6M_\odot$, and from Salasnich et al. 2000,
for $8-12 M_\odot$), for the TP-AGB phase (Marigo \& Girardi 2007;
  dotted curves), and for post-AGB evolution (Bloecker 1995b; dashed curves), 
shown in the plane of absolute $I$-band
AB magnitude, $M_I$, versus $V-I$ and $B-I$ AB colours. 
Models are labeled according
to their ZAMS masses, in Solar units, with smaller labels for the
  post-AGB tracks. The $3M_\odot$ post-AGB model 
is one which, among the models presented by Bloecker (1995b), has
a post-AGB mass of $0.625 M_\odot$. The grey areas show the regions
of parameter space with luminosities greater than 
 our absolute magnitude
 limits in $B$, $V$, and $I$, 
for stars at the positions of the SNe in NGC~1316 
(assuming a distance of 19~Mpc and no extinction).} 
\label{tracks}
\end{figure*}

Figure~\ref{tracks} shows, in a color-absolute-magnitude diagram,
various Solar-metallicity stellar evolution tracks.
In the range $3-6 M_\odot$ ZAMS mass, the tracks are
from Girardi et al. (2000), from the main sequence to the onset
of thermal pulsations on the AGB branch. The models shown 
include convective overshooting. Recent models by Marigo \& Girardi
(2007) for the thermal-pulsation stage on the AGB 
are shown for $3-5 M_\odot$ ZAMS masses.
For $8-12 M_\odot$ stars, the models are by Salasnich et al. (2000)
between the ZAMS and carbon core ignition, assuming Solar composition.
The dashed curves show post-AGB tracks by Bloecker (1995b)
for stars of ZAMS masses $3-7 M_\odot$. Bolometric luminosities
and effective temperatures of the models were converted to
broad-band AB magnitudes by integrating the surface fluxes
 in stellar-atmosphere model
spectra over the appropriate bandpasses. The solar-metallicity 
stellar atmospheres
for a grid of temperatures and surface gravities were interpolated 
to the temperatures and gravities of the stellar evolution tracks. The
stellar atmospheres are from the BaSeL 2.2 database (Lejeuene et
al. 1997, 1998; Westera et al. 2002), and from Kurucz (1992) for
temperatures above $4000 K$. 
For the post-AGB models,
monochromatic luminosities were obtained assuming a blackbody spectrum, an
approximation which is accurate to a few percent for these hot stars
and the optical bands considered here.
Models are labeled according
to their ZAMS masses, in Solar units. The grey-shaded areas show the regions
of parameter space with luminosities greater than
 our absolute magnitude
 limits in $B$, $V$, and $I$, assuming no extinction, as listed
in Table~1. Adoption of the plausible extinctions listed in Table~1 
would shift the grey region upward by 0.1~mag, and to the left by
0.15~mag (left panel) and 0.3~mag (right panel).

We see that our upper limits on the stellar luminosities at the sites
of the two SNe, do probe some of the parameter space, even if small,
of the late-time evolution of intermediate-mass stars. Specifically,
according to the models shown,
stars with ZAMS masses $\ga 6M_\odot$ reach luminosities higher than
our limits at the end of their AGB phases. Stars with initial masses
$\ga 9 M_\odot$ spend much of the time from the first ascent to the
red-giant branch and onward at luminosities above the limits.
Similarly,
post-AGB stars of ZAMS masses $\ga 4 M_\odot$, at the beginning of
their constant luminosity path to high temperatures on the HR diagram
(which results in a monotonic decrease in optical luminosity, see     
Fig.~\ref{tracks}), may be above our upper limits. 

However, the uncertainties that affect models of the later stages of
stellar evolution, particularly uncertainties due to convective overshooting
and the details of mixing and mass loss (e.g., Herwig 2005), 
make the conclusions based on comparison
to such models tentative. 
Indeed, Pauldrach et al. (2004), when modeling
the ultraviolet spectra of nine central stars of planetary nebulae (CSPNs), 
have found severe
discrepancies between the masses deduced 
for five of the stars, compared with those predicted
by the mass-luminosity relation from theoretical post-AGB stellar
models. Remarkably, the masses of these stars are close to the
Chandrasekhar limit. Conversely, Napiwotzki (2006) and Gesicki \&
Zijlstra (2007) have concluded, based, on different analysis methods, 
that CSPNs, including some of those analyzed by Pauldrach et al. (2004),
have masses of $\sim 0.6 M_\odot$, similar to most WDs, and with
mass-luminosity relations consistent with post-AGB model predictions. 
In any case, in standard SD models of SNe-Ia, complex binary
evolution precedes a SN-Ia explosion, and hence using single-star models
 to predict the
colours and the luminosities of the donor stars could be
inaccurate. And, if SNe-Ia  evolve from single stars within some
initial mass and metallicity ranges, which have somehow (e.g. via mass loss,
or previous binary interactions) lost their hydrogen envelopes, then the
stellar tracks for normal AGB evolution that we have used are unlikely    
to provide reliable predictions for the properties of such
progenitors. It should also be kept in mind
 that AGB stars, which are prodigious dust producers during their
thermal-pulse stages, are likely enshrouded in 
dust shells at the end of their evolution, 
and the exact point at which 
they become visible again on the post-AGN branch is uncertain
(e.g., Bloecker 1995a,b). 

In summary, we have searched pre-explosion images of the sites of two 
SNe-Ia for potential luminous stellar progenitors, with null results.
As always is the case in such studies, our null result does not 
conclusively rule out any of the scenarios or mass ranges for SN-Ia 
progenitors. Under any of the possibile scenarios of 
strong obscuration by dust (by 
unresolved clumps, or by a screen that is behind the bulk of the
stellar mass of the galaxy, or by a circumstellar shell), 
more luminous progenitors could be hidden. Furthermore, 
model predictions for the later
stages of stellar evolution are still quite uncertain, all the more so 
if binary interactions take place. Finally,
SNe 2006dd and 2006mr may have, by chance, belonged to the tardy SN-Ia
population, in which case
one would not expect them to be associated with a young population. 

Nevertheless, our study shows that deep HST observations of nearby SN
hosts can graze some interesting regions of stellar
parameter space that may be relevant for young-population SN-Ia
progenitors. Future SNe-Ia will explode also in more nearby galaxies
that have HST imagery. This will improve the statistics of such searches,
will make accessible lower stellar luminosities,
and will perhaps eventually reveal an actual progenitor. In view of the
many unknowns behind SN-Ia formation, such observational studies may
provide valuable clues. 
   



\begin{thebibliography}{}
\bibitem[Altavilla et al.(2004)]{2004MNRAS.349.1344A} Altavilla, G., et 
al.\ 2004, \mnras, 349, 1344 
\bibitem[Badenes et al.(2006)]{2006ApJ...645.1373B} Badenes, C., Borkowski, 
K.~J., Hughes, J.~P., Hwang, U., \& Bravo, E.\ 2006, \apj, 645, 1373 
\bibitem[Badenes et al.(2007)]{2007ApJ...662..472B} Badenes, C., Hughes, 
J.~P., Bravo, E., \& Langer, N.\ 2007, \apj, 662, 472 
\bibitem[Blair et al.(2007)]{2007ApJ...662..998B} Blair, W.~P., Ghavamian, 
P., Long, K.~S., Williams, B.~J., Borkowski, K.~J., Reynolds, S.~P., \& 
Sankrit, R.\ 2007, \apj, 662, 998 
\bibitem[Bloecker(1995a)]{1995A&A...297..727B} Bloecker, T.\ 1995a, \aap, 
297, 727 
\bibitem[Bloecker(1995b)]{1995A&A...299..755B} Bloecker, T.\ 1995b, \aap, 
299, 755 
\bibitem[Cassisi et al.(1998)]{1998ApJ...496..376C} Cassisi, S., Iben, 
I.~J., \& Tornambe, A.\ 1998, \apj, 496, 376 
\bibitem[Dahlen et al.(2004)]{2004ApJ...613..189D} Dahlen, T., et al.\ 
2004, \apj, 613, 189 
\bibitem[Dahlen et al.(2008)]{2008arXiv0803.1130D} Dahlen, T., Strolger, 
L.-G., \& Riess, A.~G.\ 2008, ArXiv e-prints, 803, arXiv:0803.1130 
\bibitem[Della Valle et al.(2005)]{2005ApJ...629..750D} Della Valle, M., 
Panagia, N., Padovani, P., Cappellaro, E., Mannucci, F., \& Turatto, M.\ 
2005, \apj, 629, 750 
\bibitem[Dilday et al.(2008)]{2008arXiv0801.3297D} Dilday, B., et al.\ 
2008, ArXiv e-prints, 801, arXiv:0801.3297 
\bibitem[Ekers et al.(1983)]{1983A&A...127..361E} Ekers, R.~D., Goss, 
W.~M., Wellington, K.~J., Bosma, A., Smith, R.~M., \& Schweizer, F.\ 1983, 
\aap, 127, 361 
\bibitem[Feldmeier et al.(2007)]{2007ApJ...657...76F} Feldmeier, J.~J., 
Jacoby, G.~H., \& Phillips, M.~M.\ 2007, \apj, 657, 76 
\bibitem[Gal-Yam et al.(2007)]{2007ApJ...656..372G} Gal-Yam, A., et al.\ 
2007, \apj, 656, 372 
\bibitem[Geldzahler \& Fomalont (1984)]{geldzahler84}
Geldzahler, B. J.,  \& Fomalont, E. B.,  1984, AJ, 89, 1650
\bibitem[Gesicki \& Zijlstra(2007)]{2007A&A...467L..29G} Gesicki, K., \& 
Zijlstra, A.~A.\ 2007, \aap, 467, L29 
\bibitem[Girardi et al.(2000)]{2000AAS..141..371G} Girardi, L., Bressan, 
A., Bertelli, G., \& Chiosi, C.\ 2000, \aaps, 141, 371
\bibitem[Goudfrooij et al.(2001a)]{2001MNRAS.322..643G} Goudfrooij, P., 
Mack, J., Kissler-Patig, M., Meylan, G., \& Minniti, D.\ 2001a, \mnras, 322, 
643 
\bibitem[Goudfrooij et al.(2001b)]{2001MNRAS.328..237G} Goudfrooij, P., 
Alonso, M.~V., Maraston, C., \& Minniti, D.\ 2001b, \mnras, 328, 237 
\bibitem[Goudfrooij et al.(2004)]{2004ApJ...613L.121G} Goudfrooij, P., 
Gilmore, D., Whitmore, B.~C., \& Schweizer, F.\ 2004, \apjl, 613, L121 
\bibitem[Greggio \& Renzini(1983)]{1983AA...118..217G} Greggio, L., \& 
Renzini, A.\ 1983, \aap, 118, 217 
\bibitem[Greggio(2005)]{2005A&A...441.1055G} Greggio, L.\ 2005, \aap, 441, 
105
\bibitem[Hachisu et al.(1996)]{1996ApJ...470L..97H} Hachisu, I., Kato, M., 
\& Nomoto, K.\ 1996, \apjl, 470, L97 
\bibitem[Hamuy et al.(1991)]{1991AJ....102..208H} Hamuy, M., Phillips, 
M.~M., Maza, J., Wischnjewsky, M., Uomoto, A., Landolt, A.~U., \& Khatwani, 
R.\ 1991, \aj, 102, 208
 \bibitem[Hamuy et al.(1994)]{1994AJ....108.2226H} Hamuy, M., et al.\ 1994, 
\aj, 108, 2226 
\bibitem[Hamuy et al.(2006)]{2006PASP..118....2H} Hamuy, M., et al.\ 2006, 
\pasp, 118, 2 
\bibitem[Hendry et al.(2006)]{2006MNRAS.369.1303H} Hendry, M.~A., et al.\ 
2006, \mnras, 369, 1303 
\bibitem[Herwig(2005)]{2005ARA&A..43..435H} Herwig, F.\ 2005, \araa, 43, 
435 
\bibitem[Hillebrandt \& Niemeyer(2000)]{2000ARAA..38..191H} Hillebrandt, 
W., \& Niemeyer, J.~C.\ 2000, \araa, 38, 191 
\bibitem[Iben \& Tutukov(1984)]{1984ApJS...54..335I} Iben, I., Jr., \& 
Tutukov, A.~V.\ 1984, \apjs, 54, 335
 \bibitem[Ihara et al.(2007)]{2007arXiv0706.3259I} Ihara, Y., Ozaki, J., 
Doi, M., Shigeyama, T., Kashikawa, N., Komiyama, Y., \& Hattori, T.\ 2007, 
PASJ, 59, 811
\bibitem[Immler et al.(2006)]{2006CBET..554....1I} Immler, S., Milne, P., 
\& Brown, P.~J.\ 2006, Central Bureau Electronic Telegrams, 554, 1 
\bibitem[Jensen et al.(2003)]{2003ApJ...583..712J} Jensen, J.~B., Tonry, 
J.~L., Barris, B.~J., Thompson, R.~I., Liu, M.~C., Rieke, M.~J., Ajhar, 
E.~A., \& Blakeslee, J.~P.\ 2003, \apj, 583, 712 
\bibitem[Jha et al.(2007)]{2007ApJ...659..122J} Jha, S., Riess, A.~G., \& 
Kirshner, R.~P.\ 2007, \apj, 659, 122 
\bibitem[Kim \& Fabbiano(2003)]{2003ApJ...586..826K} Kim, D.-W., \& 
Fabbiano, G.\ 2003, \apj, 586, 826 
\bibitem[Kobayashi et al.(1998)]{1998ApJ...503L.155K} Kobayashi, C., 
Tsujimoto, T., Nomoto, K., Hachisu, I., \& Kato, M.\ 1998, \apjl, 503,
L155 
\bibitem[Kobayashi \& Nomoto(2007)]{2008arXiv0801.0215K} Kobayashi, C., \& 
Nomoto, K.\ 2008, ApJ, submitted, arXiv:0801.0215 
\bibitem[Kurucz(1992)]{1992IAUS..149..225K} Kurucz, R.~1992, in
 Proceedings of the IAU Symposium 149, Eds. B. Barbuy and A. Renzini
 (Kluwer Academic Publishers, Dordrecht), p.225
\bibitem[Kuznetsova et al.(2008)]{2008ApJ...673..981K} Kuznetsova, N., et 
al.\ 2008, \apj, 673, 981 
\bibitem[Lejeune et 
al.(1997)]{1997A&AS..125..229L} Lejeune, T., Cuisinier, F., \& Buser,
  R.\ 1997, \aaps, 125, 229 
\bibitem[Lejeune et 
al.(1998)]{1998A&AS..130...65L} Lejeune, T., Cuisinier, F., \& Buser, R.\ 1998, \aaps, 130, 65 
\bibitem[Leonard(2007)]{2007ApJ...670.1275L} Leonard, D.~C.\ 2007, \apj, 
670, 1275 
\bibitem[Marigo \& Girardi(2007)]{2007A&A...469..239M} Marigo, P., \& 
Girardi, L.\ 2007, \aap, 469, 239 
\bibitem[Mattila et al.(2005)]{2005A&A...443..649M} Mattila, S., Lundqvist, 
P., Sollerman, J., Kozma, C., Baron, E., Fransson, C., Leibundgut, B., \& 
Nomoto, K.\ 2005, \aap, 443, 649 
\bibitem[{{Mannucci} {et~al.}(2006){Mannucci}, {Della Valle}, \&
  {Panagia}}]{Mannucci_06}
{Mannucci}, F., {Della Valle}, M., \& {Panagia}, N. 2006, \mnras, 370, 773
\bibitem[{{Mannucci} {et~al.}(2005){Mannucci}, {Della Valle}, {Panagia},
  {Cappellaro}, {Cresci}, {Maiolino}, {Petrosian}, \& {Turatto}}]{Mannucci_05}
{Mannucci}, F., {Della Valle}, M., {Panagia}, N., {Cappellaro}, E., {Cresci},
  G., {Maiolino}, R., {Petrosian}, A., \& {Turatto}, M. 2005, \aap,
  433, 807
\bibitem[Mannucci et al.(2008)]{2008MNRAS.383.1121M} Mannucci, F., Maoz, 
D., Sharon, K., Botticella, M.~T., Della Valle, M., Gal-Yam, A., 
\& Panagia, N.\ 2008, \mnras, 383, 1121 
\bibitem[Mannucci et al.(2007)]{2007MNRAS.377.1229M} Mannucci, F., Della 
Valle, M., \& Panagia, N.\ 2007, \mnras, 377, 1229
\bibitem[Maoz(2008)]{2008MNRAS.384..267M} Maoz, D.\ 2008, \mnras, 384, 267 
\bibitem[Maund \& Smartt(2005)]{2005MNRAS.360..288M} Maund, J.~R., \& 
Smartt, S.~J.\ 2005, \mnras, 360, 288
\bibitem[Mazzali et al.(2007)]{2007Sci...315..825M} Mazzali, P.~A., 
R{\"o}pke, F.~K., Benetti, S., \& Hillebrandt, W.\ 2007, Science, 315,
825 
\bibitem[Monard(2006)]{2006CBET..553....1M} Monard, L.~A.~G.\ 2006, Central 
Bureau Electronic Telegrams, 553, 1 
\bibitem[Monard \& Folatelli(2006)]{2006CBET..723....1M} Monard, L.~A.~G., 
\& Folatelli, G.\ 2006, Central Bureau Electronic Telegrams, 723, 1 
\bibitem[Morrell et al.(2006)]{2006CBET..564....1M} Morrell, N., Folatelli, 
G., Barba, R., \& Arias, J.\ 2006, Central Bureau Electronic Telegrams, 
564, 1 
\bibitem[Napiwotzki et al.(2004)]{2004ASPC..318..402N} Napiwotzki, R., et 
al.\ 2004, Spectroscopically and Spatially Resolving the Components of the 
Close Binary Stars, 318, 402 
\bibitem[Napiwotzki(2006)]{2006A&A...451L..27N} Napiwotzki, R.\ 2006, \aap, 
451, L27 
\bibitem[Nelemans et al.(2005)]{2005AA...440.1087N} Nelemans, G., et al.\ 
2005, \aap, 440, 1087 
\bibitem[Nomoto(1982)]{1982ApJ...253..798N} Nomoto, K.\ 1982, \apj, 253, 
798 
\bibitem[Nomoto \& Iben(1985)]{1985ApJ...297..531N} Nomoto, K., \&
  Iben, I., Jr.\ 1985, \apj, 297, 531 
\bibitem[]{} Patat, F. et al. 2007, Science, 317, 924 
\bibitem[Pauldrach et al.(2004)]{2004A&A...419.1111P} Pauldrach, A.~W.~A., 
Hoffmann, T.~L., \& M{\'e}ndez, R.~H.\ 2004, \aap, 419, 1111 
\bibitem{}Pavlovsky, C., et al. 2005, "ACS Data Handbook", 
Version 4.0, (Baltimore: STScI)
\bibitem[Phillips et al.(2006)]{2006CBET..729....1P} Phillips, M.~M., 
Folatelli, G., Contreras, C., \& Morrell, N.\ 2006, Central Bureau 
Electronic Telegrams, 729, 1 
\bibitem[Piersanti et al.(1999)]{1999ApJ...521L..59P} Piersanti, L., 
Cassisi, S., Iben, I.~J., \& Tornamb{\'e}, A.\ 1999, \apjl, 521, L59
\bibitem[Piersanti et al.(2000)]{2000ApJ...535..932P} Piersanti, L., 
Cassisi, S., Iben, I.~J., \& Tornamb{\'e}, A.\ 2000, \apj, 535, 932 
\bibitem[Piersanti et al.(2003)]{2003ApJ...583..885P} Piersanti, L., 
Gagliardi, S., Iben, I.~J., \& Tornamb{\'e}, A.\ 2003, \apj, 583, 885 
\bibitem[Poznanski et al.(2002)]{2002PASP..114..833P} Poznanski, D., 
Gal-Yam, A., Maoz, D., Filippenko, A.~V., Leonard, D.~C., \& Matheson, T.\ 
2002, \pasp, 114, 833 
\bibitem[Poznanski et al.(2007)]{2007MNRAS.382.1169P} Poznanski, D., et 
al.\ 2007, \mnras, 382, 1169 
\bibitem[Prieto et al.(2007)]{2007arXiv0707.0690P} Prieto, J.~L., Stanek, 
K.~Z., \& Beacom, J.~F.\ 2007, ApJ, submitted, arXiv:0707.0690 
\bibitem[Reynolds et al.(2007)]{2007ApJ...668L.135R} Reynolds, S.~P., 
Borkowski, K.~J., Hwang, U., Hughes, J.~P., Badenes, C., Laming, J.~M., \& 
Blondin, J.~M.\ 2007, \apjl, 668, L135 
\bibitem[Ruiz-Lapuente et al.(2004)]{2004Natur.431.1069R} Ruiz-Lapuente, 
P., et al.\ 2004, \nat, 431, 1069 
\bibitem[Salasnich et al.(2000)]{2000A&A...361.1023S} Salasnich, B., 
Girardi, L., Weiss, A., \& Chiosi, C.\ 2000, \aap, 361, 1023 
\bibitem[Salvo et al.(2006)]{2006CBET..557....1S} Salvo, M., Blackman, J., 
Schmidt, B., \& Bessell, M.\ 2006, Central Bureau Electronic Telegrams, 
557, 1 
\bibitem[{{Scannapieco} \& {Bildsten}(2005)}]{Scannapieco_05}
{Scannapieco}, E., \& {Bildsten}, L. 2005, \apjl, 629, L85
\bibitem[Schlegel et al.(1998)]{1998ApJ...500..525S} Schlegel, D.~J., 
Finkbeiner, D.~P., \& Davis, M.\ 1998, \apj, 500, 525 
\bibitem[Schweizer(1980)]{1980ApJ...237..303S} Schweizer, F.\ 1980, \apj, 
237, 303 
\bibitem[Sharon et al.(2007)]{2007ApJ...660.1165S} Sharon, K., Gal-Yam, A., 
Maoz, D., Filippenko, A.~V., \& Guhathakurta, P.\ 2007, \apj, 660, 1165 
\bibitem[Simon et al.(2007)]{2007ApJ...671L..25S} Simon, J.~D., et al.\ 
2007, \apjl, 671, L25 
\bibitem[Sullivan et al.(2006)]{2006ApJ...648..868S} Sullivan, M., et al.\ 
2006, \apj, 648, 868 
\bibitem[Tout(2005)]{2005ASPC..330..279T} Tout, C.~A.\ 2005, in The 
Astrophysics of Cataclysmic Variables and Related Objects , 330, 279 
\bibitem[Van Dyk et al.(1999)]{1999AJ....118.2331V} Van Dyk, S.~D., Peng, 
C.~Y., Barth, A.~J., \& Filippenko, A.~V.\ 1999, \aj, 118, 2331 
\bibitem[]{} Waldman, R., Yungelson, L, and Barkat, Z. 2007, astro-ph/0710.3911
\bibitem[Webbink(1984)]{1984ApJ...277..355W} Webbink, R.~F.\ 1984, \apj, 
277, 355 
\bibitem[West et al.(1987)]{1987A&A...177L...1W} West, R.~M., Lauberts, A., 
Schuster, H.-E., \& Jorgensen, H.~E.\ 1987, \aap, 177, L1 
\bibitem[Westera et 
al.(2002)]{2002A&A...381..524W} Westera, P., Lejeune, T., Buser, R., Cuisinier, F., \& Bruzual, G.\ 2002, \aap, 381, 524 
\bibitem[Whelan \& Iben(1973)]{1973ApJ...186.1007W} Whelan, J., \& Iben, 
I.~J.\ 1973, \apj, 186, 1007 
\bibitem[White \& Malin(1987)]{1987Natur.327...36W} White, G.~L., \& Malin, 
D.~F.\ 1987, \nat, 327, 36 


\end{thebibliography}
\end{document}